\documentclass[pra, singlecolumn, preprintnumbers, amsmath, amssymb]{revtex4-2}

\usepackage[dvips]{graphicx}
\usepackage{srcltx}
\usepackage{color}
\usepackage{bm,upgreek}
\usepackage{amsmath,mathtools}
\usepackage{amsfonts}
\usepackage{amssymb}
\usepackage{mathrsfs}
\usepackage{boxedminipage}
\usepackage{framed}
\usepackage{bbm}
\usepackage{natbib}
\usepackage[ugly]{nicefrac}
\usepackage{mathtools}
\usepackage{enumitem}
\usepackage{multirow}
\usepackage{slashed}
\usepackage{tikz-feynman}

\def\del#1{\color{grey}\sout{#1}~\color{black}}

\def\acomm#1#2{{\big\{#1,#2\big\}}}           
\def\mod2#1{{\big|#1\big|^2}}                 
\def\av#1{{\langle#1\rangle}}                 
\def\-{\!-\!}                               
\def\+{\!+\!}                               
\def\={\;=\;}                               

\def\a{\vec{a}}

\def\k{\vec{k}}

\def\p{\vec{p}}

\def\x{\vec{x}}

\def\A{\vec{A}}

\def\0{\vec{0}}

\def\d{\mathrm{d}}                          
\def\intd#1{\int\d#1\;}                     
\def\intd2#1{\int\d^2#1\;}                  
\def\intd3#1{\int\d^3#1\;}                  
\def\grad{\boldsymbol{\nabla}}              
\def\del{\partial}                          
\def\vec#1{\mathbf{#1}}                     
\def\-{\!-\!}                               
\def\+{\!+\!}                               

\def\beq{\begin{equation}}                  
\def\eeq{\end{equation}}                    

\def\aslash{\slashed{a}}
\def\bslash{\slashed{b}}
\def\pslash{\slashed{p}}
\def\kslash{\slashed{k}}
\def\Aslash{\slashed{A}}
\def\epsilonslash{\slashed{\epsilon}}

\begin{document}


\title{Laser-induced phase shift of swift electrons}

\author{Christian Dwyer}
\email{christian.dwyer@dectris.com}
\affiliation{DECTRIS Ltd., Baden-Daettwil, Switzerland}
\affiliation{Physics, School of Science, RMIT University, Melbourne, Victoria 3001, Australia}

\author{David M. Paganin}
\email{david.paganin@monash.edu}
\affiliation{School of Physics and Astronomy, Monash University, Clayton, Victoria 3800, Australia}

\begin{abstract}

We revisit the calculation of the phase shift experienced by swift electrons on passing through the electromagnetic field of a laser. Such phase shifts are now utilized in the form of `laser phase plates' in transmission electron microscopes (TEMs), for example. We calculate the phase shift using three different methods, namely, perturbation theory applied to the Dirac equation, the Volkov solution to the Klein-Gordon equation, and the relativistic Hamilton-Jacobi equation. We find that all three methods are in agreement, and that the calculated phase shift is independent of the relative orientation of the electron and laser beams. The agreement between the quantum and classical theories is explained. The duality between the phase shifts experienced by electrons, photons and neutrons is described. Our Lorentz invariant result for the phase shift differs from certain results published in the literature.

\end{abstract}

\maketitle 

\nopagebreak

\section{Introduction}
\label{sec:introduction}

Recent years have witnessed significant advances in the realization of ``laser phase plates" which capitalize on the fact that swift electrons ($\sim100$~keV) which pass through a laser's electromagnetic field experience a phase shift. Such phase plates have been used in a transmission electron microscope (TEM) to improve the image contrast of weakly-scattering samples, especially biological samples, via the Zernike phase-contrast mechanism. The interaction of electrons with a laser is also applied in other areas, notably free electron lasers and laser-based electron accelerators.

Several works in the literature present calculations and applications of the electron phase shift. The following is an incomplete list of relevant articles relating to laser phase plates in the TEM in recent decades:

\begin{itemize}

\item Kaplan and Pokrovsky~(2005) \cite{KaplanPokrovsky2005} --- These authors use numerical integration of the fully-relativistic Lorentz equation. Note, however, that this work is not aimed at TEM and it does not include a calculation of the phase shift.

\item M\"{u}ller et al.~(2010) \cite{Muller2010} --- These authors use quantum electrodynamics perturbation theory. They obtain a result for the phase shift which is anisotropic, in the sense that the phase shift depends on the relative orientation of the electron and laser 3-momenta.

\item Axelrod et al.~(2020) \cite{Axelrod2020} --- These authors begin with the nonrelativistic Lagrangian in the electron's rest frame and then boost to the laboratory frame. For an electron with energy $E$ moving along the $z$ axis, they obtain the following expression for the phase shift (in natural units)
\beq\nonumber -\frac{e^2}{2E} \int \d t \left[ (\A_\perp - \grad_\perp\Lambda)^2 + (A_z - \del_z \Lambda)^2/\gamma^2 \right], \eeq
where $A_\perp=(A_x,A_y)$ and $A_z$ are the Cartesian components of the electromagnetic vector potential, $\Lambda$ is a gauge function, $\gamma$ is the Lorentz factor, and the integral is evaluated along the free-electron trajectory in the laboratory frame as a function of the laboratory time $t$.

\item García de Abajo and Konečná~(2021) \cite{GarciadeAbajoKonecna2021} --- These authors use a quasi-first-order expansion of the Dirac equation in the photon's 4-momentum $k$. They obtain for the phase shift (in natural units)
\beq\nonumber -\frac{e^2}{2E} \int \d t\left[ \A_\perp^2 + A_z^2 / \gamma^2\right], \eeq 
which is similar to the expression of Axelrod et al.~(2020) \cite{Axelrod2020}, however, the integrand does not contain a gauge function. The factor of $\gamma$ implies an anisotropic phase shift for relativistic electrons. For example, in the case of a linearly-polarized laser with 3-momentum perpendicular to the electron beam, a different phase shift is obtained depending on whether the  polarization is parallel or perpendicular to the electron beam.

\item Uesugi and Kozawa~(2025) \cite{UesugiKozawa2025} --- These authors use the nonrelativistic result obtained by setting $\gamma = 1$ and $\Lambda = 0$ in the expressions above. The different results published in the literature agree in nonrelativistic limit. However, electrons in the TEM are relativistic.

\item Petrov et al.~(2026) \cite{Petrov2026} --- This work reports a recent implementation of laser phase plates. These authors also quote the nonrelativistic result.

\end{itemize}

In this work, we revisit the calculation of the laser-induced electron phase shift. Since several different results appear in the literature, and they cannot all be correct, it is timely to revisit the calculation. Moreover, we have performed the calculation in three different ways in an attempt to resolve the discrepancies in the literature. Our calculated laser-induced swift-electron phase shift is shown to be Lorentz invariant, which is a natural requirement since it is directly proportional to the difference in action between two spacetime points along the electron trajectory. We are concerned specifically with the case where the electron's initial velocity is relativistic, but the laser intensity is ``nonrelativistic," that is, it does not cause the electron's path to differ significantly from the field-free case. 

Our units and most of our notation follows Bjorken and Drell~(1964) \cite{BjorkenDrell1964}, except that, unlike those authors, we always regard $e$ as a positive number, namely, the magnitude of the electron's charge. In these natural units, $\hbar = c = \epsilon_0 = 1$, the reduced Compton wavelength is $1/m = 0.386$~pm, the electron rest energy is $m=511$~keV, the fine-structure constant is $\alpha = e^2/4\pi \approx 1/137$, and the unit of charge is $\sqrt{\epsilon_0\hbar c} = 5.291\times10^{-19}$~C. We work in the laboratory frame unless stated otherwise, and we use the radiation gauge.

\section{Dirac equation using perturbation theory}
\label{sec:phase shift of Dirac electron}

\begin{figure}[t!]
\begin{center}
\scalebox{1.0}{\includegraphics{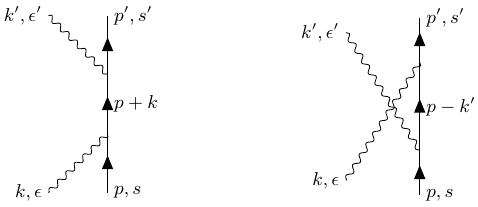}}
\end{center}
\caption{\label{fig:stimulated Compton Feynman diagrams}Feynman diagrams for $S^{(2)}_{fi}$.}
\end{figure}

In this section, we derive the laser-induced electron phase shift using second-order perturbation theory applied to the Dirac equation. Corresponding to the Feynman diagrams in Fig.~\ref{fig:stimulated Compton Feynman diagrams}, wherein time runs from bottom to top, the relevant (second-order) scattering matrix element consists of two terms:
\beq\label{eq:Compton scattering matrix element} \begin{split} S^{(2)}_{fi} = &-ie^2 \int \d^4x\d^4y\, \bar\psi_f(x)\Aslash(x;k') S_F(x-y)\slashed{A}(y;k)\psi_i(y) \\
&- ie^2 \int \d^4x\d^4y\, \bar\psi_f(x) \Aslash(x;k) S_F(x-y)\slashed{A}(y;k')\psi_i(y). \end{split}\eeq
Here, $\psi_i$ and $\psi_f$ are the initial and final Dirac 4-component electron wave functions, $S_F$ is the relativistic free-electron propagator, and the factors of $\Aslash\equiv A^\mu\gamma_\mu$ come from the 4-potential $A^\mu$ of each photon, where $\mu = 0,\dots,3$ is a spacetime index, $\gamma^\mu$ are the Dirac matrices, and the summation convention is used over the repeated spacetime index $\mu$. Note the notation $\bar\psi_f\equiv \psi_f^\dag\gamma^0$, where $\psi_f^\dag$ denotes the adjoint. The integrations are performed over all spacetime points $x$ and $y$.

We stipulate that the electron has initial and final 4-momenta $p$ and $p'$, and initial and final spins $s$ and $s'$. The electron's initial wave function has the form
\beq \psi_i(x) = \sqrt{\frac{m}{EV}}u(p,s)e^{-ip\cdot x},\eeq
where $E \equiv p^0$ is the initial energy, which we assume to be positive, $u(p,s)$ is a 4-component Dirac spinor, and $V$ is the volume for box normalization. To obtain the electron's final wave function $\psi_f$ we replace $p,s$ with $p',s'$.

The ingoing and outgoing photons have 4-momenta $k$ and $k'$, respectively. For the ingoing photon, $A$ is given by
\beq A^\mu(x;k) = \frac{\epsilon^\mu}{\sqrt{2\omega V}}(e^{-ik\cdot x} + e^{+ik\cdot x}),\eeq
where $k$ is the 4-momentum for which $k\cdot k =0$, $\omega=k^0$ is the photon's energy, and $\epsilon$ is the polarization 4-vector. The 4-momentum and polarization obey the condition $k\cdot\epsilon=0$. Here the box normalization is such that, at any given time, the total energy in the box is equal to the photon's energy $\omega$. For the outgoing photon, we replace $k,\epsilon$ with $k',\epsilon'$, which obey $k' \cdot k'=0$ and $k'\cdot\epsilon'=0$. In the laboratory frame, the radiation gauge is used, so that $\epsilon$ and $\epsilon'$ have only spatial components, i.e., $\epsilon=(0,\boldsymbol{\epsilon})$ with $|\boldsymbol{\epsilon}|=1$ and $\epsilon'=(0,\boldsymbol{\epsilon}')$ with $|\boldsymbol{\epsilon}'|=1$, which imply $\k\cdot\boldsymbol{\epsilon}=0$ and $\k'\cdot\boldsymbol{\epsilon}'=0$. 

For initial and final states as specified above, the spacetime integrals are Fourier transforms, and so we obtain
\beq  S^{(2)}_{fi} = -\frac{ie^2 m (2\pi)^4\delta^{(4)} }{\sqrt{EE'} \sqrt{2\omega2\omega'} V^2} \bar u(p',s') \left(\epsilonslash' \frac{\pslash+\kslash+m}{(p+k)^2-m^2} \epsilonslash + \epsilonslash \frac{\pslash-\kslash'+m}{(p-k')^2-m^2} \epsilonslash' \right)u(p,s),\eeq
where $\delta^{(4)}$ is our shorthand for $\delta^{(4)}(p'+k'-p-k)$, which specifies conservation of the 4-momentum. For any two slashed symbols $\aslash$ and $\bslash$, we have $ \aslash\bslash = 2a\cdot b - \bslash\aslash$. Therefore, since $k\cdot\epsilon=k'\cdot\epsilon'=0$, we have $\epsilonslash\kslash = -\kslash\epsilonslash$ and $\epsilonslash'\kslash' = -\kslash'\epsilonslash'$. Using these commutation relations, and simplifying the denominators, we can obtain
\beq\begin{split}  &S^{(2)}_{fi} = -\frac{ie^2 m(2\pi)^4\delta^{(4)} }{\sqrt{EE'} \sqrt{2\omega2\omega'} V^2} \\
&\times \bar u(p',s') \bigg(\epsilonslash' \frac{2p\cdot\epsilon - \epsilonslash(\pslash +\kslash -m)}{2p\cdot k}
 + \epsilonslash \frac{2p\cdot\epsilon' - \epsilonslash'(\pslash -\kslash' -m)}{- 2p\cdot k'}\bigg)u(p,s). \end{split}\eeq
Then using $(\pslash-m)u(p,s) = 0$, we get
\beq S^{(2)}_{fi} = -\frac{ie^2 m(2\pi)^4\delta^{(4)}}{\sqrt{EE'} \sqrt{2\omega2\omega'} V^2} \bar u(p',s') \left(\epsilonslash' \frac{ 2p\cdot\epsilon - \epsilonslash \kslash}{2p\cdot k}  + \epsilonslash \frac{ 2p\cdot\epsilon' + \epsilonslash' \kslash'}{-2p\cdot k'}\right)u(p,s). \eeq

At this stage of the calculation, we stipulate that the scattering process ultimately leaves the electron and photon states unchanged, that is, we set $p'=p$, $s'=s$, $k'=k$ and $\epsilon'=\epsilon$. The justification for this stipulation is given below. Thus, we now write the matrix element as $S^{(2)}_{ii}$. The equality $\epsilon=\epsilon'$ implies $\epsilonslash'\epsilonslash = \epsilonslash\epsilonslash' = -1$, and we obtain
\beq \label{eq:Compton matrix element intermediate} S^{(2)}_{ii} = -\frac{ie^2 (2\pi)^4\delta^{(4)}(0)}{2 E \omega V^2} \frac{m}{p\cdot k} \bar u(p,s) \kslash u(p,s), \eeq
where $\delta^{(4)}(0)$ is now an infinite factor which will be addressed later. To evaluate the spinor product in Eq.~\eqref{eq:Compton matrix element intermediate}, we write the 4-component Dirac spinors $u$ in terms of the 2-component Pauli spinors $u_s, u_s'$:
\beq \bar u(p,s') = \sqrt{\frac{E+m}{2m}}\begin{pmatrix} \bar u_{s'} & - \frac{p}{E+m} \bar u_{s'} \sigma^3 \end{pmatrix},
\quad u(p,s) = \sqrt{\frac{E+m}{2m}}\begin{pmatrix} u_{s} \\ \frac{p}{E+m} \sigma^3 u_{s} \end{pmatrix}, \eeq
where we momentarily allow for the possibility of a spin flip $s'\neq s$, and for any slashed symbol $\aslash$
\beq \aslash = a^0 \gamma^0 - \a\cdot\boldsymbol{\gamma} = \begin{pmatrix} a^0 & - \a\cdot\boldsymbol{\sigma} \\ \a\cdot\boldsymbol{\sigma} & -a^0   \end{pmatrix} . \eeq
We obtain for the spinor product
\beq \label{eq:spinor product} \bar u(p,s') \aslash u(p,s) = \frac{1}{2m} \bar u_{s'} \left( 2E a^0 - p \acomm{\a\cdot\boldsymbol{\sigma}}{\sigma^3}   \right) u_{s} = \frac{p\cdot a}{m}\bar u_{s'} u_{s} = \frac{p\cdot a}{m} \delta_{s' s}, \eeq
where the second equality follows from the anticommutator result $\acomm{\a\cdot\boldsymbol{\sigma}}{\sigma^3} = 2 a^3$, and we see that spin flips are excluded. Using the above result for the spinor product (with $s'=s$), Eq.~\eqref{eq:Compton matrix element intermediate} simplifies to
\beq \label{eq:Compton matrix element} S^{(2)}_{ii} = -\frac{ie^2 (2\pi)^4\delta^{(4)}(0)}{2 E \omega V^2}. \eeq
This result holds for all orientations of the electron and photon beams. 

In a manner typical of particle-scattering calculations, we can replace the infinite factor $(2\pi)^4 \delta^{(4)}(0)$ with $TV$, where $T$ is the infinite time during which the scattering process has been allowed to take place, and $V$ is the infinite box volume. Following M\" uller et al.\ (2010) \cite{Muller2010}, we can then replace $T$ with a finite time $\Delta t$. The remaining uncanceled factor $1/V$ can be interpreted as the photon number density, which can be replaced by a finite photon number density $\rho$. We obtain
\beq  S^{(2)}_{ii} = -\frac{ie^2 \rho\Delta t}{2 E \omega}. \eeq

To see that the matrix element $S^{(2)}_{ii}$ really does correspond to a phase shift of the electron's wave function, consider the Born series for the exact matrix element: $S_{fi} = \delta_{fi} + S^{(1)}_{fi} + S^{(2)}_{fi} + \cdots$. In the present case, the first-order contribution $S_{fi}^{(1)}$ vanishes because such an interaction between an electron and a photon cannot conserve 4-momentum. Setting $i=f$, we get $S_{ii} \approx 1 + S_{ii}^{(2)} = 1 + i\phi\approx e^{i\phi}$. Hence $S_{ii}^{(2)}$ does indeed correspond to a phase shift, given by
\beq \label{eq:Compton phase shift} \phi =  -\frac{e^2 \rho\Delta t}{2 E \omega} = -\frac{e^2 \rho\Delta z}{2 p k} = -\frac{\alpha \rho \lambda_\gamma \lambda_e \Delta z}{2\pi}, \eeq
where the second equality follows from $p=\beta E$ and $\Delta z = \beta\Delta t$, $\beta$ being the electron's speed relative to the speed of light, and the third equality uses the electron and photon wavelengths $\lambda_e = 2\pi/p$ and $\lambda_\gamma = 2\pi/k$, respectively. Equation~\eqref{eq:Compton phase shift} agrees with Refs.~\cite{Muller2010, Axelrod2020, GarciadeAbajoKonecna2021} in the nonrelativistic limit $\gamma\rightarrow1$. Away from that limit, it agrees with the expression of Axelrod et al.~(2020) \cite{Axelrod2020}, but not those in Refs.~\cite{Muller2010, GarciadeAbajoKonecna2021}.

Regarding our stipulation that the electron and photon states are ultimately unchanged, note that the theory certainly does not demand this. Rather, the stipulation is justified if we envisage that the density of photons with 4-momenta $k$ is high enough that the virtual electron undergoes \emph{stimulated} emission. The situation is very similar to conventional stimulated emission in a laser. As is well-known from the theory of the quantized electromagnetic field \cite{Loudon2000}, the presence of $n_k$ photons in the same state $k,\epsilon$ leads to a factor of $\sqrt{n_k}$ in a photon absorption matrix element, and a factor of $\sqrt{n_k+1}$ in a photon emission matrix element. With reference to Fig.~\ref{fig:stimulated Compton Feynman diagrams}, the diagram on the left is multiplied by $n_k$, while the diagram on the right is multiplied by $n_k+1$. If $n_k$ is very large, then the difference between $n_k$ and $n_k+1$ is negligible, and we effectively obtain an additional very large factor $n_k$ for our matrix element $S^{(2)}_{ii}$. Further, if, in the limit $V\rightarrow\infty$, we also allow $n_k\rightarrow\infty$, such that $n_k/V$ is constant, then this procedure is consistent with our earlier replacement of $1/V$ by a finite photon number density $\rho=n_k/V$. Finally, a matrix element for scattering into final states different from the initial ones will be multiplied by factor of $\sqrt{n_k}$, which is negligible compared to $n_k$. Therefore, the matrix element $S^{(2)}_{ii}$ dominates.

\section{Klein-Gordon equation using Volkov solution}
\label{sec:phase shift from Volkov solution}

For our second calculation of the laser-induced electron phase shift, we use the well-known Volkov solutions \cite{Volkov1935, JentschuraAdkins2022} to the Dirac and Klein-Gordon equations. While the Dirac and Klein-Gordon equations give the same result for the phase shift, here we choose the Klein-Gordon equation as a point of distinction and, moreover, the analysis is simpler and none of the important physics is missed.

The Klein-Gordon equation describing the motion of a spinless electron moving in a $\tau$-parametrized 4-potential $A$ (to be defined below) can be written in the form
\beq ((i\nabla)^2 + 2eA\cdot i\nabla + e^2A^2 - m^2)\psi = 0, \eeq
where $i\nabla$ is the 4-momentum operator and $\psi$ is a scalar wave function of position and time. The 4-potential $A$ is considered to be a function of the parameter $\tau = k\cdot x$. Moreover, $A$ is assumed to vanish in the limits $\tau\rightarrow\pm\infty$. For such 4-potentials, the solutions to the Klein-Gordon equation are the Volkov solutions
\beq \psi_p^\pm = \exp\left( -ip\cdot x + \frac{i}{2p\cdot k} \int_{\mp\infty}^{k\cdot x} \d\tau \left(2eA\cdot p + e^2A^2 \right)  \right). \eeq
The 4-vector $p$ which parametrizes the Volkov solutions $\psi_p^+$ and $\psi_p^-$ is the momentum of the electron in the asymptotic past and asymptotic future, respectively. For convenience, let us use the Klein-Gordon equation to define an interaction potential $V$:
\beq ((i\nabla)^2  - m^2)\psi =  -(2eA\cdot i\nabla + e^2A^2)\psi = V \psi \quad\implies\quad V = -2eA\cdot p - e^2A^2 . \eeq
Now the Volkov solutions can be written more succinctly in the form
\beq \psi_p^\pm = \exp\left( -ip\cdot x - \frac{i}{2p\cdot k}  \int_{\mp\infty}^{k\cdot x} \d\tau \, V \right). \eeq

As discussed by Meyer~(1970) \cite{Meyer1970}, and Dawson and Fried~(1970) \cite{DawsonFried1970}, the boundary conditions of the Volkov solutions can require special attention. Specifically, since the 4-potential $A(\tau)$ remains finite along the spacetime directions $\pm k$ (where $\tau=0$), it does not vanish at all spatial positions in the asymptotic past or asymptotic future. This implies that the Volkov solutions do not reduce to free-space solutions at those times. Using Volkov wave packets, rather than the Volkov solutions themselves, avoids such difficulties because the electron in the asymptotic past and future can be effectively confined to the regions where $A$ vanishes. This allows us to maintain that $A$ is strictly a function of the parameter $\tau$, which is a fundamental characteristic of the Volkov solutions, and enables a considerably more elegant mathematical analysis than introducing a time window function, such as $e^{-\epsilon |t|}$, say, which would break the $\tau$ parametrization. However, as we shall see, a correction to the result obtained for the phase shift is required to compensate for the fact that, within the $\tau$ parametrization, the wave packet's interaction time inadvertently becomes dependent on $p\cdot k$.

Following Meyer~(1970) \cite{Meyer1970}, we consider a normalized wave packet of positive-energy free-space solutions to the Klein-Gordon equation, given by
\beq\begin{split} f(x) &= \int\frac{\d^4 p}{(2\pi)^3}\, \delta(p^2-m^2)\sqrt{p_0+|p_0|} f(\p) 
e^{ -ip\cdot x }\\
&= \int\frac{\d^4 p}{(2\pi)^3}\, \frac{\delta(p^0 - E) + \delta(p^0+E)}{2E} \sqrt{p_0+|p_0|} f(\p) 
e^{ -ip\cdot x }\\
&= \int\frac{\d^3 \p}{(2\pi)^3}\, \frac{f(\p)}{\sqrt{2 E}} e^{ -iE t + i\p\cdot\x }. \end{split}\eeq
Let us define a complete set of such wave packets by way of the orthonormality relation
\beq\begin{split} \int\d^3\x\,\bar f_\alpha(x) \overleftrightarrow{i\del_0} f_\beta(x) &= \int \frac{\d^4 p}{(2\pi)^3}\, \delta(p^2-m^2)(p_0-|p_0|) \bar f_\alpha(\p) f_\beta(\p)\\
&= \int \frac{\d^3\p}{(2\pi)^3}\, \bar f_\alpha(\p) f_\beta(\p)\\
&= \delta_{\alpha\beta}. \end{split}\eeq

We will use the 3-momentum amplitudes $f_\alpha(\p)$ defined above to construct Volkov wave packets. However, before doing so, let us first analyze the dispersion properties of a Gaussian wave packet in the Klein-Gordon theory, which will help to solidify the previous statements regarding the confinement of the electron to the regions where the 4-potential $A$ vanishes in the asymptotic past and future. To that end, let us define a Gaussian wave packet at the initial time $t=0$
\beq f(0,\x) = \left(\frac{1}{2\pi\sigma^2} \right)^{3/4}\exp(-\x^2/4\sigma^2 + i\p\cdot\x),\eeq
where $\p$ is the central 3-momentum and $\sigma$ is the initial width. The Fourier transform is
\beq \int\d^3 \x\, f(0,\x)e^{-i\p'\cdot\x} = \left(8\pi\sigma^2\right)^{3/4}\exp(-\sigma^2(\p'-\p)^2) = \frac{f(\p')}{\sqrt{2E'}}.\eeq 
We can propagate the initial Gaussian wave packet over a time $t$ to examine its behavior:
\beq f(t,\x) = \int\frac{\d^3\p'}{(2\pi)^3}\, \frac{f(\p')}{\sqrt{2E'}}e^{-iE't+i\p'\cdot\x}.\eeq
In calculating the inverse Fourier transform above, if the spread in momentum space is sufficiently small, then we can expand $E'$ in the exponent to second order
\beq E' \approx E + \vec{v}_g\cdot (\p'-\p) + \frac{1}{2!}(\p'-\p)^T\alpha (\p'-\p),\eeq
where $\vec{v}_g = \p/E$ is the group velocity and $\alpha = \mathrm{diag}(\alpha_x, \alpha_y, \alpha_z) = \mathrm{diag}(1,1,\gamma)/\gamma m$ is the dispersion tensor. Then, using well-known results on the inverse Fourier transform of a Gaussian function, we can obtain
\beq \label{eq:Gaussian wave packet} f(x) = \left(\frac{\sigma^2}{2\pi} \right)^{3/4} \frac{e^{-ip\cdot x}}{\sqrt{\det M}}\exp\left( -(\x-\vec{v}_g t)^T M^{-1} (\x-\vec{v}_g t)\right). \eeq
Here $M = \sigma^2 + i\alpha t/2$ is a diagonal matrix and $M^{-1}$ is its inverse, given by 
\beq M^{-1} = \mathrm{diag}\left(\frac{1-i\alpha_x t/2\sigma^2}{4\sigma_x^2(t)}, \frac{1-i\alpha_y t/2\sigma^2}{4\sigma_y^2(t)}, \frac{1-i\alpha_z t/2\sigma^2}{4\sigma_z^2(t)} \right), \eeq
where $\sigma_x^2(t) = \sigma^2 + \alpha^2_x t^2 / 4 \sigma^2$, and similarly for $\sigma_y^2(t)$ and $\sigma_z^2(t)$, are time-dependent variances. Let us write the latter variances in the form $\sigma^2(t) = \sigma^2 + \alpha^2 t^2 / 4 \sigma^2$. 

Apart from the time-dependent phase factors, $f(x)$ in Eq.~\eqref{eq:Gaussian wave packet} is a Gaussian wave packet moving at the group velocity $\vec{v}_g$ and spreading out according to $\sigma(t)$. For long times $t\rightarrow\infty$, we obtain $\sigma(t) \simeq  \alpha t/2\sigma$. For the wave packet and the $A$ field to remain sufficiently well-separated in the asymptotic future, we require $\sigma(t) \ll t$, or $2\sigma \gg 1/\gamma m$, that is, the initial localization should not approach the reduced Compton wavelength. This requirement is easily fulfilled. 

With the aid of the 3-momentum amplitudes $f_\alpha(\p)$, we define the corresponding Volkov wave packets, given by
\beq\begin{split} F_\alpha^\pm(x) &= \int\frac{\d^4 p}{(2\pi)^3}\, \delta(p^2-m^2)\sqrt{p_0+|p_0|} f_\alpha(\p) \psi_p^\pm(x) \\
&=\int\frac{\d^3 \p}{(2\pi)^3}\,\frac{f_\alpha(\p)}{\sqrt{2E}} \psi_{E,\p}^\pm(x) \\
&=\int\frac{\d^3 \p}{(2\pi)^3}\,\frac{f_\alpha(\p)}{\sqrt{2E}} 
\exp\left( -iE t + i\p\cdot\x - \frac{i}{2(E\omega-\p\cdot\k)}\int_{\mp\infty}^{k\cdot x} \d\tau \, V \right). \end{split}\eeq
We refer to $F_\alpha^+(x)$ and $F_\alpha^-(x)$ as ``ingoing" and ``outgoing" Volkov wave packets, since they reduce to the free wave packet $f_\alpha(x)$ in the asymptotic past and the asymptotic future, respectively. The ingoing Volkov wave packets $F_\alpha^+$ have the same orthonormality relation as the free wave packets $f_\alpha(x)$
\beq \int \d^3\x\, \bar F^+_\alpha(x) \overleftrightarrow{i\del_0} F_\beta^+(x) = \int \frac{\d^3\p}{(2\pi)^3}\, \bar f_\alpha(\p) f_\beta(\p)\\
= \delta_{\alpha\beta}. \eeq
Likewise for the outgoing Volkov wave packets $F_\alpha^-$.

With the Volkov wave packets defined as above, our goal now is to calculate the phase shift experienced by an ingoing Volkov wave packet $F_\beta^+$ in the asymptotic future. As in Section \ref{sec:phase shift of Dirac electron}, we proceed to calculate elements of the scattering matrix $S$. Here, the required matrix element is that between the ingoing Volkov wave packet $F_\beta^+$ and a state that reduces to a free wave packet $f_\alpha$ in the asymptotic future, namely, the outgoing Volkov wave packet $F_\alpha^-$. The matrix element is
\beq S_{\alpha\beta} =  \int \d^3\x\, \bar F^-_\alpha(x) \overleftrightarrow{i\del_0} F_\beta^+(x) =  \lim_{t\rightarrow\infty} \int \d^3\x\, \bar f_\alpha(x) \overleftrightarrow{i\del_0} F_\beta^+(x). \eeq
The second equality follows because $t$ can have any value and $F_\alpha^-(x)$ reduces to $f_\alpha(x)$ in the asymptotic future. Following a careful calculation whose details we shall omit, the result is (Meyer, 1970)~\cite{Meyer1970}
\beq S_{\alpha\beta} = \int \frac{\d^3 \p}{(2\pi)^3}\,  \bar f_\alpha(\p) f_\beta(\p) \exp\left( - \frac{i}{2p\cdot k}  \int_{-\infty}^{+\infty} \d\tau \, V \right). \eeq
The form of $S_{\alpha\beta}$ is similar to the orthonormality relation for $F_\alpha^+$ and $F_\beta^+$, except that the integrand of $S_{\alpha\beta}$ contains a phase factor. Choosing each of the $f_\alpha(\p)$'s to define a narrow band centered at some $\p_\alpha$, the phase factor can be taken outside the integral, to obtain
\beq S_{\alpha\beta} = \exp\left( - \frac{i}{2p\cdot k} \int_{-\infty}^{+\infty} \d\tau \, V \right) \delta_{\alpha\beta}. \eeq 
Setting $\alpha = \beta$, and assuming that the contribution from the oscillatory term $-2eA\cdot p$ in $V$ averages to zero, we obtain that the phase shift experienced by an ingoing Volkov wave packet $F_\alpha^+$ is 
\beq\label{eq:phase shift from Volkov} \phi = \frac{1}{2p\cdot k} \int_{-\infty}^{+\infty} \d\tau \,e^2A^2. \eeq
Equation~\eqref{eq:phase shift from Volkov} is valid for any $\tau$-parametrized 4-potential $A(\tau)$. While we omit the details, Eq.~\eqref{eq:phase shift from Volkov} can also be obtained by adopting a $\tau$-parametrized 3-potential $\A(\tau)$ in the expression of Axelrod et al.~(2020) \cite{Axelrod2020} quoted in Section \ref{sec:introduction}. Thus, Eq.~\eqref{eq:phase shift from Volkov} agrees with Axelrod et al.~(2020) \cite{Axelrod2020}.

To connect with the results of Section \ref{sec:phase shift of Dirac electron}, we take the 4-potential to be that of a linearly polarized electromagnetic wave in the radiation gauge
\beq\label{eq:Volkov 4-potential} A(\tau) =  \frac{a(\tau)\epsilon}{\sqrt{2\omega V}} (e^{-i\tau} + e^{+i\tau} ) = \frac{a(\tau)\epsilon}{\sqrt{2\omega V}}2  \cos \tau, \eeq
where $a(\tau)$ is a ``window" function which restricts the 4-potential to a finite $\tau$ interval. With this definition, the integral in Eq.~\eqref{eq:phase shift from Volkov} is finite and independent of $p$ and $k$.

When the electron and laser 3-momenta are orthogonal, we have $p\cdot k = E\omega$, and the phase shift is given by
\beq \phi = \frac{1}{2p\cdot k} \int_{-\infty}^{+\infty}\d\tau\, e^2 A^2 = - \frac{e^2}{E \omega^2 V} \int_{-\infty}^{+\infty}\d\tau\, \cos^2 \tau =  - \frac{e^2}{E \omega^2 V} \frac{2\pi N}{2}, \eeq
where $N$ is the number of $2\pi$ intervals integrated over. We can write $2\pi N / \omega = T$, where $T$ is the interaction time. Then, as in Section \ref{sec:phase shift of Dirac electron}, we can replace $T$ with a finite time $\Delta t$, and replace $1/V$ with a finite photon density $\rho$, to obtain for the phase shift
\beq \phi = - \frac{e^2 \rho \Delta t}{2 E \omega}, \eeq
which agrees with our previous result. 

\begin{figure}[t!]
\begin{center}
\scalebox{0.68}{\includegraphics{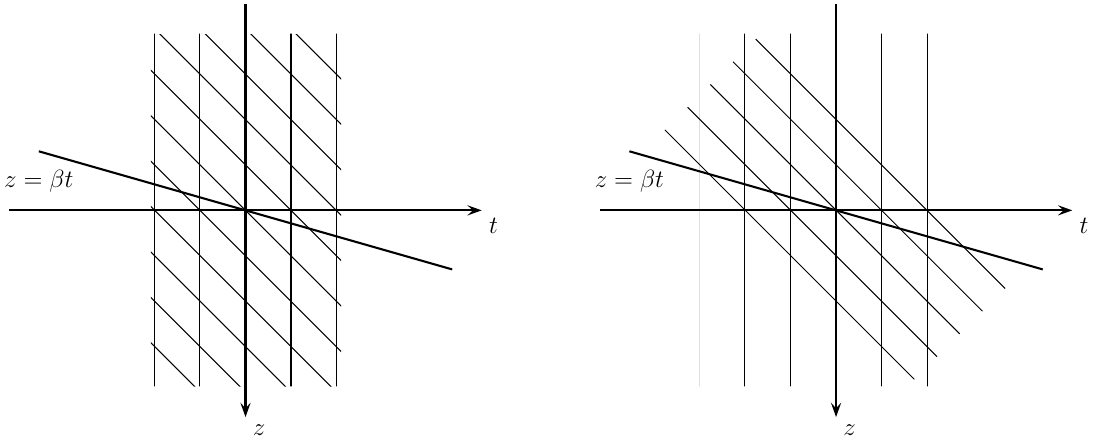}}
\caption{\label{fig:Volkov phase shift} Graphical explanation of different boundary conditions for the radiation field $A$. The vertical lines represent planes of constant $\tau$ for $A$ propagating perpendicular to $z$, while the lines oriented at 45 degrees are those for $A$ propagating parallel to $z$. The straight line $z=\beta t$ represents an electron wave packet trajectory (or a classical trajectory) which begins at a point in the asymptotic past before the $A$ field is present, passes through the field at finite times, and ends at a point in the asymptotic future when the field is again absent. Left: the field $A$ is confined to a specific time interval regardless of its orientation. Right: the field $A$ is confined to a specific $\tau$ interval which, in the parallel case, implies an interaction time which is longer by a factor $(1-\beta)^{-1}$. In the latter case, a multiplicative correction factor of $1-\beta$ must be applied to obtain the phase shift $\phi$ corresponding to the specific time interval.}
\end{center}
\end{figure}

When the electron and laser 3-momenta are parallel, the integral over $\tau$ is unchanged, but we have $p\cdot k = E\omega(1-\beta)$. At first sight then, it appears that a greater phase shift is obtained for the parallel case. However, consideration of Fig.~\ref{fig:Volkov phase shift} shows that, in confining the 4-potential to a specific $\tau$ interval, the interaction time in the parallel case is longer by a factor $(1-\beta)^{-1}$, which must be corrected. The correction cancels the factor $(1-\beta)^{-1}$, leading to the same phase shift as the perpendicular case. These results agree with those obtained in Section \ref{sec:phase shift of Dirac electron}.

\section{Relativistic Hamilton-Jacobi equation}
\label{sec:phase shift from classical mechanics}

For our third calculation of the laser-induced electron phase shift, we use the relativistic Hamilton-Jacobi equation 
\beq \left( - \del_t S + eA^0 \right)^2 - \left(\grad S(x) + e \A \right)^2 - m^2 = 0, \eeq
where $S$ is the action of an electron moving in the 4-potential $A$, $\grad S + eA$ is the electron's kinetic momentum and $-\del_t S$ is its energy. In relativistic notation, the Hamilton-Jacobi equation reads
\beq (-\nabla S + e A)^2 - m^2 = 0. \eeq
Following Sarachik and Schappert~(1970) \cite{SarachikSchappert1970}, in the case of a $\tau$-parametrized 4-potential $A(\tau)$, the ansatz is 
\beq\label{eq:action ansatz} S(x) = - p\cdot x + \phi(\tau), \eeq 
where $p$ is a constant 4-vector to be determined from the boundary conditions, and the function $\phi(\tau)$ is determined by satisfying the Hamilton-Jacobi equation. The form of $\phi(\tau)$ can be deduced to be
\beq \phi(\tau) = \frac{1}{2 p\cdot k} \int_{\tau_0}^\tau\d\tau' \left(p^2 - m^2 + 2 e p\cdot A + e^2 A^2 \right). \eeq
As is implied by our notation, we find that the constant 4-vector $p$ can be set equal to the electron's 4-momentum in the asymptotic past. In the expression for $\phi(\tau)$ above, the vanishing term $p^2 - m^2$ is retained as it yields a nonvanishing result when calculating the electron's trajectory in Section \ref{sec:classicality of the phase shift}.

For the action in the asymptotic future, we can drop the vanishing term $p^2 - m^2$ as well as the integral of $p\cdot A$ which averages to zero, to obtain
\beq S(x) = - p\cdot x +  \frac{1}{2 p\cdot k} \int_{-\infty}^{+\infty}\d\tau\, e^2 A^2 , \eeq
which contains the same phase offset as obtained previously.

We note that, if we are given an action $S$ obeying the relativistic Hamilton-Jacobi equation, then the wave function $e^{iS}$ satisfies the Klein-Gordon equation \cite{MotzSelzer1964}. In this sense, it can be argued that our derivation in this section is not distinct from that in Section \ref{sec:phase shift from Volkov solution}. On the other hand, that classical mechanics predicts the same result as quantum mechanics is noteworthy and will be discussed further below.

\section{Lorentz invariance of the phase shift}
\label{Lorentz invariance of the phase shift}

Physically, any phase shift is a Lorentz invariant because it dictates the change in an interference pattern, which is a concept that can be quantified in a manner that is independent of the observer's inertial frame. Mathematically, our swift-electron phase shift is intrinsically Lorentz invariant since (in natural units) it is equal to the difference in action between two spacetime points along the electron trajectory [cf.~Eqn.~\eqref{eq:action ansatz}]. Therefore, we consider the Lorentz invariance of our result
\beq \phi = - \frac{e^2\rho\Delta t}{2E\omega}. \eeq
Specifically, let us consider performing a Lorentz boost which takes us from the laboratory frame into the electron's rest frame denoted by barred coordinates:
\beq\begin{split} \bar t &= \gamma (t - \beta z),\\
\bar x &= x,\\
\bar y &= y,\\
\bar z &= \gamma (z - \beta t). \end{split} \eeq
In the rest frame, we have $\bar E = m$ and $\Delta \bar t = \Delta t / \gamma$. If the electron and laser 3-momenta are perpendicular, we have $\bar\omega = \gamma\omega$ and $\bar\rho = \gamma\rho$, and hence we obtain that the phase shift is Lorentz invariant. If the electron and laser 3-momenta are parallel, we have $\bar\omega = \gamma(1-\beta)\omega$ and $\bar\rho = \gamma(1-\beta)\rho$, and again we obtain that the phase shift is Lorentz invariant. Note that in the latter case, $\bar\rho$ is not determined by a simple Lorentz contraction. 

We note that the expression of Axelrod et al.~(2020) \cite{Axelrod2020} is Lorentz invariant. On the other hand, in the perpendicular case, and with the linear polarization parallel to the electron beam, the expression for the phase shift published by García de Abajo and Konečná~(2021) \cite{GarciadeAbajoKonecna2021} contains an additional factor $1/\gamma^2$, which, on transforming to the electron's rest frame, becomes unity, and so their result is not Lorentz invariant. Similarly, in the parallel case, the result published by M\"{u}ller et al.~\cite{Muller2010} is not Lorentz invariant.

\section{Classicality of the phase shift}
\label{sec:classicality of the phase shift}

It is remarkable that the quantum and classical theories give exactly the same result for the laser-induced electron phase shift. Whenever quantum and classical mechanics agree, it is wise to suspect that ``there is a harmonic oscillator lurking somewhere''~\cite{Moodie2010}. 

Indeed, in the inertial frame in which the electron's average 3-momentum vanishes, the laser's electromagnetic field causes the electron to undergo motion which, when parametrized by $\tau$, is harmonic \cite{SarachikSchappert1970}. To see this, we consider the electron's classical trajectory as given by the  Hamilton-Jacobi theory of Section \ref{sec:phase shift from classical mechanics}. The trajectory is obtained by equating $-\nabla_p S$ to the electron's initial coordinate $x(\tau_0)$, which gives the electron's spacetime coordinates in the laboratory frame as a function of $\tau$ 
\beq\label{eq:electron trajectory vs tau} x(\tau) = x_0 + \frac{1}{p\cdot k} \int_{\tau_0}^\tau\d\tau' \left( p + e A \right) - \frac{k}{2 (p\cdot k)^2} \int_{\tau_0}^\tau\d\tau' \left(2 e p\cdot A + e^2 A^2\right). \eeq
The average 3-momentum arises from two terms, namely, the $p$ term in the first integral and the average value of $A^2$ in the second integral. Let us use barred coordinates for the inertial frame with vanishing average 3-momentum. We obtain from Eq.~\eqref{eq:electron trajectory vs tau}
\beq\label{eq:harmonic motion} \bar \x(\tau) = \bar \x_0 + \frac{1}{\bar p\cdot \bar k} \int_{\tau_0}^\tau\d\tau' e \bar \A  - \frac{\bar \k}{2(\bar p\cdot \bar k)^2} \int_{\tau_0}^\tau\d\tau'\, e^2\left(\bar A^2 - \av{\bar A^2}\right). \eeq
The first integral in Eq.~\eqref{eq:harmonic motion} produces motion along $\bar \A$ with frequency $\bar\omega$, while the second integral produces motion along $\bar \k$ with frequency $2\bar\omega$. The electron's total motion is a ``figure of eight'' orbit in the plane spanned by $\bar\A$ and $\bar\k$ \cite{SarachikSchappert1970}. This is our harmonic oscillator. 

Note that the presence of $\bar A$ alters the temporal coordinate too. This implies that the electron's motion as a function of $\bar t$ is more complex than Eq.~\eqref{eq:harmonic motion}. In particular, the motion as a function of $\bar t$ is not harmonic. In the case of a $\tau$-parameterized 4-potential this does not matter because the motion with respect to $\tau$ is harmonic, and so the classical phase shift will agree with the quantum one. On the other hand, if a more general form of the 4-potential is allowed, as in the work of Axelrod et al.~(2020) \cite{Axelrod2020}, then the classical phase shift must be viewed as only an approximation to the quantum one. The approximation will be good for ``nonrelativistic'' laser intensities where the motion with respect to $\bar t$ is approximately harmonic, but it will break down at higher laser intensities.

As a final remark on the classicality of the phase shift, we note that the agreement of the quantum and classical theories in the case of a $\tau$-parameterized 4-potential $A$ is also attributable to the fact that $A$ has been regarded as an external potential. If we were to go beyond such a description to include effects such as radiation reaction, we should again expect the agreement of the quantum and classical theories to break down.

\section{Dual experiments with photons and neutrons}
\label{sec:dualities}

Our focus on the phase shift of electrons passing through a laser invites a comparison with a dual experiment wherein photons traverse a film of randomly-positioned stationary electrons. Such a dual experiment can be realized by passing hard x-rays through an amorphous film of low atomic number; hard x-rays interact primarily with the electron density in the film, and not the atomic nuclei. The phase shift $\phi_\gamma$ experienced by the photons is
\beq \label{eq:photon phase shift} \phi_\gamma = -\frac{\alpha \rho_e \lambda_C \lambda_\gamma \Delta z}{2\pi},\eeq
where $\rho_e$ is the electron number density and $\lambda_C$ is the electron Compton wavelength. Equation~\eqref{eq:photon phase shift} can be derived independently of the present work by substituting Eq.~(2.134) into Eq.~(2.41) of Paganin (2006) \cite{Paganin2006}. However, Eq.~\eqref{eq:photon phase shift} can also be obtained from Eq.~\eqref{eq:Compton phase shift} as follows: Firstly, we set $\rho = 1/V$, as it was originally in Section \ref{sec:phase shift of Dirac electron}, to give the electron's phase shift due to its interaction with a single photon. Then we argue the key point that, by symmetry, exactly the same phase shift must be experienced by a photon interacting with a single electron. The electron's energy can have any value, including $E = m = 2\pi/\lambda_C$ for an electron at rest. For the interaction time, we set $\Delta t = \Delta z$, as appropriate for a photon traversing a film of thickness $\Delta z$ containing a randomly-positioned stationary electron. Finally, in analogy with the replacement made in Section \ref{sec:phase shift of Dirac electron}, we replace $1/V$ with a finite electron density $\rho_e$. The result is Eq.~\eqref{eq:photon phase shift}. It is natural that the fine structure constant $\alpha$ appears in both Eqs.~\eqref{eq:Compton phase shift} and \eqref{eq:photon phase shift} since both scenarios involve electrons and photons interacting with one another.

In a separate dual experiment, neutrons traverse a film of randomly positioned stationary nuclei.  Such a dual experiment can be realized by passing neutrons through a film composed of a nuclear isotope. The phase shift $\phi_n$ experienced by the neutrons is
\beq \label{eq:neutron phase shift} \phi_n = -\rho_n b \lambda_n \Delta z,\eeq
where $\rho_n$ is the number density of nuclei in the film, $b$ is the bound neutron scattering length, and $\lambda_n$ is the neutron wavelength. Equation~\eqref{eq:neutron phase shift}, from which a second duality is evident, can be readily obtained from the Fermi thin-slab formula \cite{FermiMarshall1947, Bacon1975, Paganin-etal2023}. The fine-structure constant is absent since the relevant interaction is now the weak force rather than the electromagnetic force (the intrinsic strength of the weak interaction is implicit in the neutron scattering length $b$).

\end{document}